\documentclass[prl,reprint,amsmath,amssymb]{revtex4-1}
\usepackage{graphicx}
\begin{document}

\title{Poles of the Scattering Matrix: An Inverse Method for Designing Photonic Resonators}

\author{Brian Slovick}\email{Corresponding author: brian.slovick@sri.com}
\author{Erik Matlin}
\affiliation{Applied Optics Laboratory, SRI International, Menlo Park, California 94025, United States}
\begin{abstract}
We develop and implement a new mathematical and computational framework for designing photonic elements with one or more high-$Q$ scattering resonances. The approach relies on solving for the poles of the scattering matrix, which mathematically amounts to minimizing the determinant of the Fredholm integral operator of the electric field with respect to the permittivity profile of the scattering element. We apply the method to design subwavelength gradient-permittivity structures with multiple scattering resonances and quality factors exceeding 500. We also find the spectral scattering cross sections are consistent with Fano lineshapes. The compact form and computational efficiency of our formalism suggest it can be a useful tool for designing Fano-resonant structures with multiple high-$Q$ resonances for applications such as frequency mixing and conversion.
\end{abstract}
\maketitle

\section{Introduction}
The promise of enhancing nonlinear optical interactions with engineered resonances is longstanding \cite{Fiedler1993,Lin2014,Lin2016,Furst2010,Ou1993,Rodriguez2007}. Compared to bulk materials, resonant structures provide larger field enhancement and longer interaction times, resulting in nonlinear processes at lower pump powers and within smaller volumes \cite{Rodriguez2007,Kauranen2012,Liu2016,Molesky2018}. In addition, the field enhancement in resonant structures removes the phase-matching requirement for harmonic generation \cite{Kauranen2012,Liu2016}. 

Of particular interest is the enhancement of harmonic generation in structures possessing resonances at both the fundamental and harmonic frequencies \cite{Fiedler1993,Lin2014,Lin2016,Furst2010,Ou1993}. In principle, the efficiency of second harmonic generation in doubly-resonant structures scales as $Q_\omega^4 Q_{2\omega}^2$ \cite{Rodriguez2007,Sitawarin2018}, where $Q_\omega$ and $Q_{2 \omega}$ are the quality factors of the resonator at the fundamental and second harmonic frequencies. Several authors have attempted to exploit this dependence using a variety of doubly-resonant structures. Early efforts employed Fabry-Perot cavities \cite{Ou1993,Berger1997}, photonic crystals \cite{Lin2014,Liscidini2006}, and whispering gallery mode resonators \cite{Fiedler1993,Bi2012,Lake2016}. However, the large modal volumes associated with these structures limit their practical use. On the other hand, doubly-resonant plasmonic structures have been shown to enhance harmonic generation with smaller modal volumes \cite{,Thyagarajan2012,Thyagarajan2013,Celebrano2015}, but the unavoidable material loss limits their efficiency \cite{Khurgin2015, Molesky2018}.

More recently, the research has shifted to the use of Mie resonance in low-loss dielectric resonators \cite{Ginn2012,Jahani2016,Kuznetsov2016}. Mie resonance has been used to enhance harmonic generation in dielectric metasurfaces \cite{Shcherbakov2014,Grinblat2016,Liu2016}. However, Mie resonators have large radiation losses and relatively small $Q$s, limiting the gains in efficiency. Dielectric structures combining Mie resonators with non-radiating elements, such as rings, can be used to obtain Fano resonances with high $Q$s and small mode volumes \cite{Limonov2017,Yang2014}. Fano resonance in Si-based dielectric metasurfaces have been used to enhance third harmonic generation \cite{Yang2015,Shorokhov2016}. However, it is unclear how these simple designs would be extended to obtain the doubly-resonant structures required to enhance harmonic generation.

The challenge of designing structures with multiple Fano resonances for harmonic generation is better suited for an inverse design approach \cite{Lin2016,Lin2017,Sitawarin2018}. Inverse methods aim to optimize an objective functional by iteratively solving Maxwell's equations for a variety of material topologies \cite{Molesky2018}. They have been used to design ring resonators \cite{Lin2017}, fibers \cite{Sitawarin2018}, and gratings \cite{Lin2016} for nonlinear frequency conversion \cite{Lin2017,Sitawarin2018}. However, the need to solve Maxwell's equations iteratively is computationally expensive and time consuming, limiting the variety of structures that can be considered.

In this Letter, we develop a new mathematical and computational framework for designing resonant scatterers that does not require iteratively solving Maxwell's equations. Instead, our algorithm is based on optimizing a single constraint condition defined by the poles of the scattering matrix, which mathematically amounts to minimizing the determinant of the integral operator of the electric field with respect to the permittivity profile of the resonator. Here we develop the approach and implement it to design several resonant structures. First, we show that a given subwavelength structure can be designed to resonate at different prescribed wavelengths, and that the spectral cross sections are consistent with Fano lineshapes with $Q$s exceeding 500. Then we show that the same structure can be designed to resonate at two prescribed wavelengths simultaneously. The compact form, computational efficiency, and generality of our approach suggest it can be a reliable inverse design tool for engineering Fano-resonant structures for enhanced light-matter interactions.

\section{Approach}
In this section we derive the condition for scattering resonances from the integral form of Maxwell's equations. The advantage of working with the integral equation is that the boundary conditions inherent to the scattering problem are implicit. The condition for scattering resonances will be derived by recasting the integral equation into matrix form and solving for poles of the electric field.

Following the notations in Refs. \cite{Potthast2001,Colton2012,Liu2019}, let $\epsilon(x)$ be the electric permittivity, $\mu=\mu_0$ be the magnetic permeability, and $\sigma(x)$ be the electric conductivity, where $\epsilon(x)=\epsilon_0$ outside a compact region $\Omega$. We also assume isotropic media, i.e. $\epsilon$, $\mu$, and $\sigma$ are scalars. With this notation, the time-harmonic Maxwell's equation for the electric field in an inhomogeneous medium is \cite{Potthast2001,Colton2012,Liu2019}
\begin{equation}
\nabla \times(\nabla \times E(x))-k^2 (1-m(x))E(x)=0,
\end{equation}
where $k=\sqrt{\epsilon_0 \mu_0} \omega$ and $\omega$ is the angular frequency, and $m(x)$ is given by
\begin{equation}
m(x)\equiv 1-\frac{1}{\epsilon_0}\left( \epsilon_r(x)+i\frac{\sigma(x)}{\omega}\right), \nonumber
\end{equation}
where $m(x)=0$ outside $\Omega$. The electric field can be written as the sum of the incident field $E_i(x)$ and the scattered field $E_s(x)$, where $E_i(x)$ is a solution to the homogeneous equation \cite{Potthast2001,Colton2012,Liu2019}
\begin{equation}
\nabla \times(\nabla \times E_i(x))-k^2 E_i(x)=0,
\end{equation}
and $E_s(x)$ satisfies the Silver-M\"{u}ller radiation condition
\begin{equation}
\lim_{|x|\rightarrow \infty} (\nabla \times E_s(x))\times x -ik|x|E_s(x)=0.
\end{equation}
The equivalent integral equation form of Eq. (1), consistent with Eqs. (2) and (3), is \cite{Colton2012}
\begin{eqnarray}
E(x)=&&E_i(x)-k^2\int_\Omega G(x-y)m(y)E(y)dy \nonumber \\
 &&-\int_\Omega \frac{1}{1-m(y)}\nabla m(y) \cdot E(y)\nabla G(x-y)dy,
\end{eqnarray}
where $G(x)$ is the Helmholtz Green's function. In three dimensions, $G(x)=\frac{e^{ik|x|}}{4\pi |x|}$ or in two dimensions, $G(x)=\frac{i}{4}H_0^{(1)}(k|x|)$, where $H_0^{(1)}(x)$ is the Hankel function of the first kind. We also make the following definitions
\begin{equation}
p^d(x)\equiv \frac{1}{1-m(x)}\frac{\partial m}{\partial x_d} (x), \quad G^d(x)\equiv \frac{\partial G}{\partial x_d}(x), \quad d=1,2,3. \nonumber
\end{equation}
To write Eq. (4) in matrix form, we discretize $\Omega$ on a uniform Cartesian grid with $n$ points per dimension and step size $h$, and write the convolution (Toeplitz) matrices as
\begin{equation}
G_{i,j}=h^3G(ih-jh), \quad G^d_{i,j}=h^3G^d(ih-jh), \nonumber
\end{equation}
where the subscripts denote discrete indices. 
With these notations, Eq. (4) can be expressed as the following matrix equation \cite{Potthast2001,Colton2012,Liu2019}
\begin{widetext}
\begin{equation}
\begin{bmatrix} E^1 \\ E^2 \\E^3 \end{bmatrix}=\begin{bmatrix} E_i^1 \\ E_i^2 \\E_i^3 \end{bmatrix} -\begin{bmatrix} k^2Gm+G^1p^1 & G^1p^2 & G^1p^3  \\  G^2p^1 & k^2Gm+G^2p^2 & G^2p^3 \\ G^3p^1 & G^3p^2 &  k^2Gm+G^3p^3 \end{bmatrix} \begin{bmatrix} E^1 \\ E^2 \\E^3 \end{bmatrix}
\end{equation}
\end{widetext}
where $E^d$ and $E_i^d$ are discrete vectors and $m$ and $p^d$ are diagonal matrices. Defining the transition matrix $T$ as 
\begin{equation}
T \equiv -\begin{bmatrix} k^2Gm+G^1p^1 & G^1p^2 & G^1p^3  \\  G^2p^1 & k^2Gm+G^2p^2 & G^2p^3 \\ G^3p^1 & G^3p^2 &  k^2Gm+G^3p^3 \end{bmatrix}, \nonumber
\end{equation}
and $E$ and $E_i$ as the vectors of all $d$ components, Eq. (5) can be solved to obtain
\begin{equation}
E=(I-T)^{-1} E_i,
\end{equation}
where $I$ is the identity matrix. Since resonances correspond to singularities of the electric field, the condition for scattering resonances leads to the requirement \cite{Landau1996}
\begin{equation}
\text{det}(I-T)=0.
\end{equation}
Equation (7) forms the basis of our inverse design approach. It can be solved using standard optimization algorithms such as the Levenberg-Marquardt method.

\section{Results}
In this section, we apply Eq. (7) to design several photonic resonators. For demonstration, we consider a two-dimensional circular structure with transversely-polarized electric field. In this case, the second integral in Eq. (4) is zero and the condition for scattering resonances in Eq. (7) simplifies to
\begin{equation}
\text{det} (I+k^2Gm)=0.
\end{equation}
First, we design a resonator element with a single high-$Q$ resonance at a wavelength of 1 $\mu$m. To ensure the retrieved permittivity can be fabricated from realistic materials, we constrain $\epsilon_r$ to be purely real and less than 16 (corresponding to Ge). We also constrain the diameter to the shortest wavelength considered (0.4 $\mu$m). This ensures the structure can be incorporated into a non-diffracting array. With these constraints, we solve Eq. (8) using the fsolve function in MATLAB, which is based on the Levenberg-Marquardt method.

Figure 1a shows a cross sectional plot of the retrieved permittivity profile. It is approximately four-point symmetric with values ranging from 4 to 12. There are also four low-permittivity regions along the polar direction. Figure 1b shows the scattering cross section for light incident along the $y$ direction. As expected, there is a high-$Q$ resonance at the design wavelength of 1 $\mu$m. The cross section also exhibits a null near the resonance peak, indicating Fano resonance behavior \cite{Limonov2017}. To confirm this, we fit the resonance to a Fano lineshape and find excellent agreement for $Q=570$. Figure 1c shows the electric field amplitude at 1 $\mu$m. The field resembles a quadrupole mode with four maxima along the polar direction and values exceeding $8,000$ relative to the incident field. The maxima appear to coincide with the regions of low permittivity. For reference, Fig. 1d shows the field away from resonance at 0.5 $\mu$m. In this case, there is no significant field enhancement.

\begin{figure}
\includegraphics[scale=0.74]{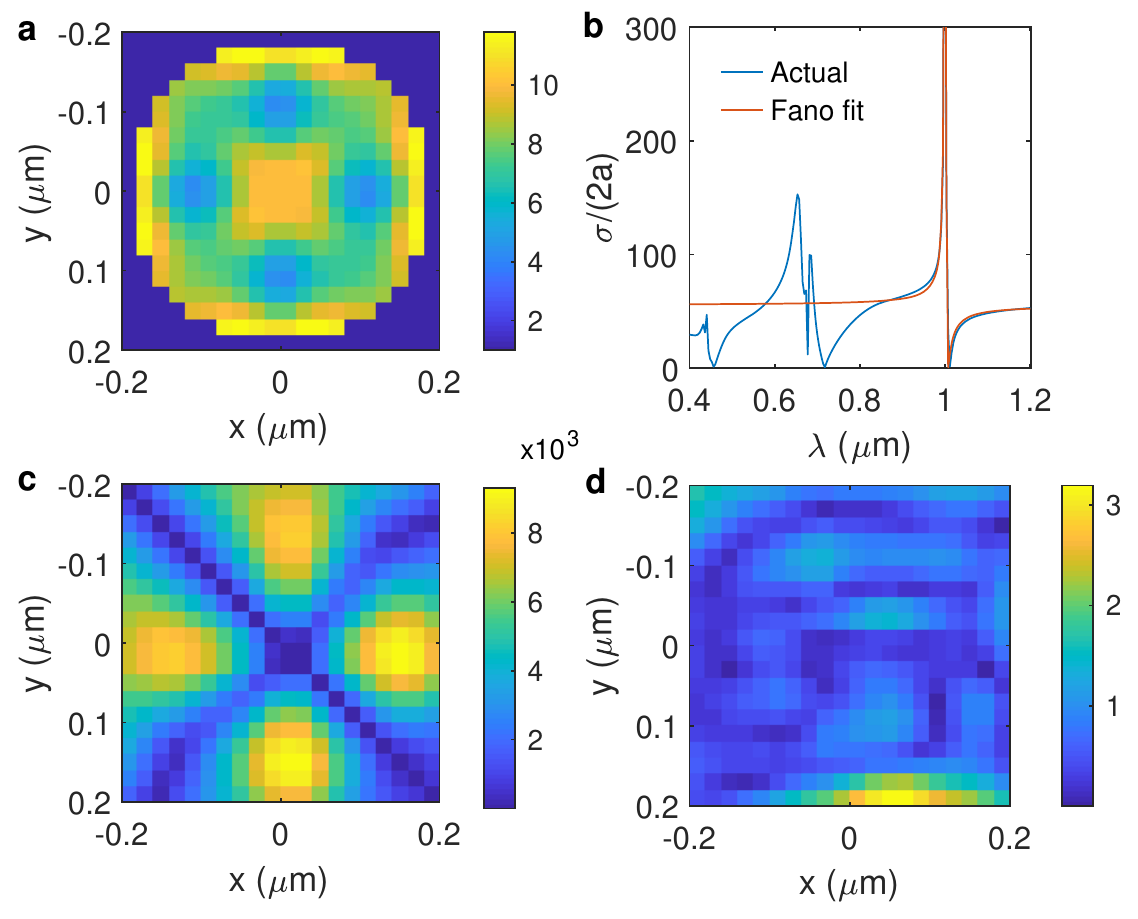}
\caption{\label{fig:epsart} Inverse design of a 0.4 $\mu$m structure with a resonance at 1 $\mu$m. (a) Permitivity profile, (b) scattering cross section, and electric field at (c) 1 $\mu$m and (d) 0.5 $\mu$m.}
\end{figure} 

Next, we apply to formalism to design a structure with the same form factor and constraints, but with a resonance at half the wavelength (0.5 $\mu$m). Figure 2a shows the retrieved permittivity profile. It ranges from 2 to 7 with eight distinct minima along the polar direction. The scattering cross section for light incident along the $y$ direction is shown in Fig. 2b. As desired, there is a high $Q$ resonance at 0.5 $\mu$m. The resonance peak fits a Fano lineshape with $Q=490$ (not shown). Figures 2c and d show the field amplitudes at 1 $\mu$m and 0.5 $\mu$m, respectively. Away from resonance at 1 $\mu$m, there is no significant field enhancement, while at resonance, the field is enhanced by more than 10$^4$. The fields are concentrated around with perimeter with eight maxima along the polar direction, consistent with an octupole mode.

\begin{figure}
\includegraphics[scale=0.74]{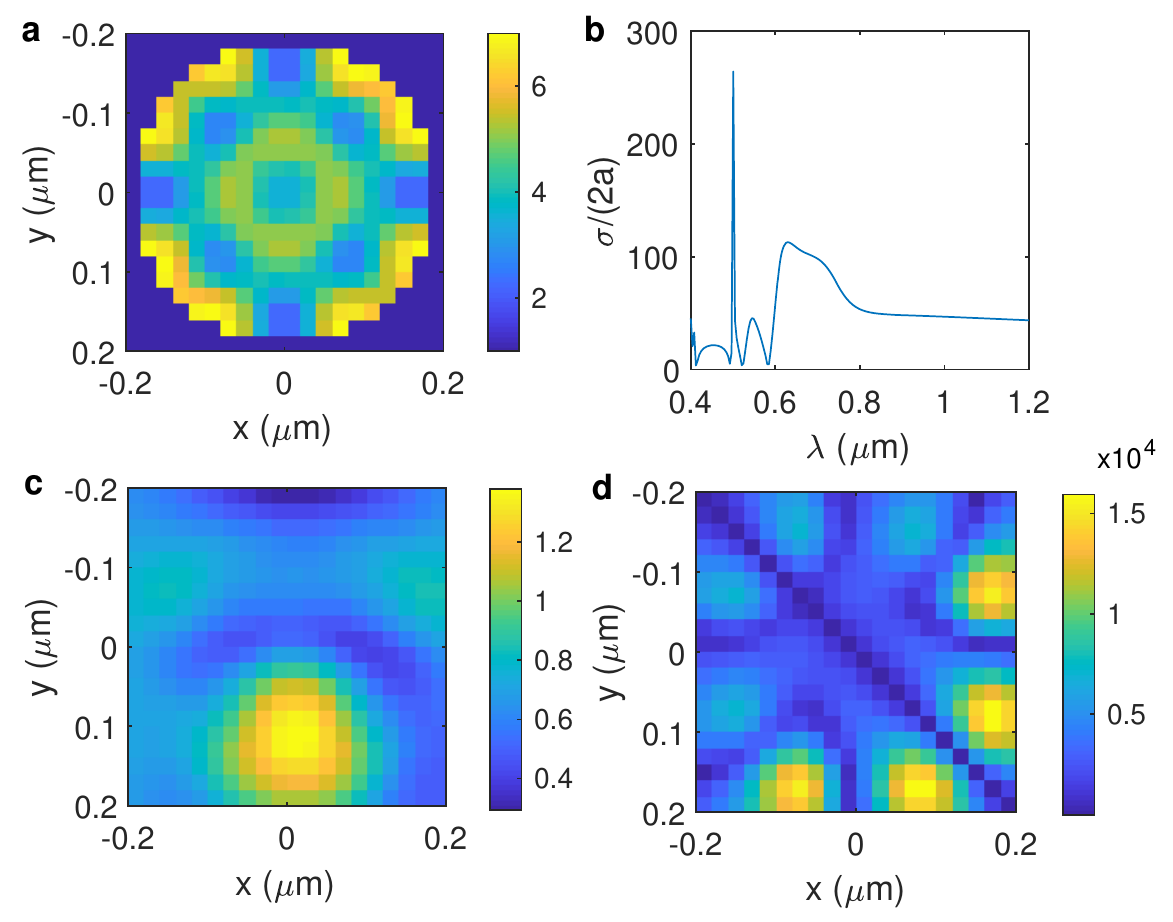}
\caption{\label{fig:epsart} Inverse design of a 0.4 $\mu$m structure with a resonance at 0.5 $\mu$m. (a) Permittivity profile, (b) scattering cross section, and electric field at (c) 1 $\mu$m and (d) 0.5 $\mu$m.}
\end{figure} 

Lastly, we apply the formalism to design a structure with the same form factor and constraints, with high-$Q$ scattering resonances at two wavelengths: 1 $\mu$m and 0.5 $\mu$m. In this case, we performed a multiobjective optimization to minimize both determinants,
\begin{equation}
\text{det} (I+k_1^2G(k_1)m)=0, \quad \text{det} (I+k_2^2G(k_2)m)=0,
\end{equation}
where the subscripts denote the two different wavelengths. Figure 3a shows the permittivity profile. It is approximately two-point symmetric with values ranging from 4 to 13. Figure 3b shows the scattering cross section for light incident along the $y$ direction. As desired, there are resonances at both 1 $\mu$m and 0.5 $\mu$m. Fitting the resonances to the Fano lineshape, we obtain $Q$s of 270 and 720 at 1 $\mu$m and 0.5 $\mu$m, respectively. Figures 3c and d show the electric field amplitudes at 1 $\mu$m and 0.5 $\mu$m, respectively. At 1 $\mu$m, the field is clearly a quadrupole mode, while at 0.5 $\mu$m the field is multimodal, with four maxima both inside and outside the structure. The field enhancement is greater than 10$^3$ at both wavelengths.

\begin{figure}
\includegraphics[scale=0.74]{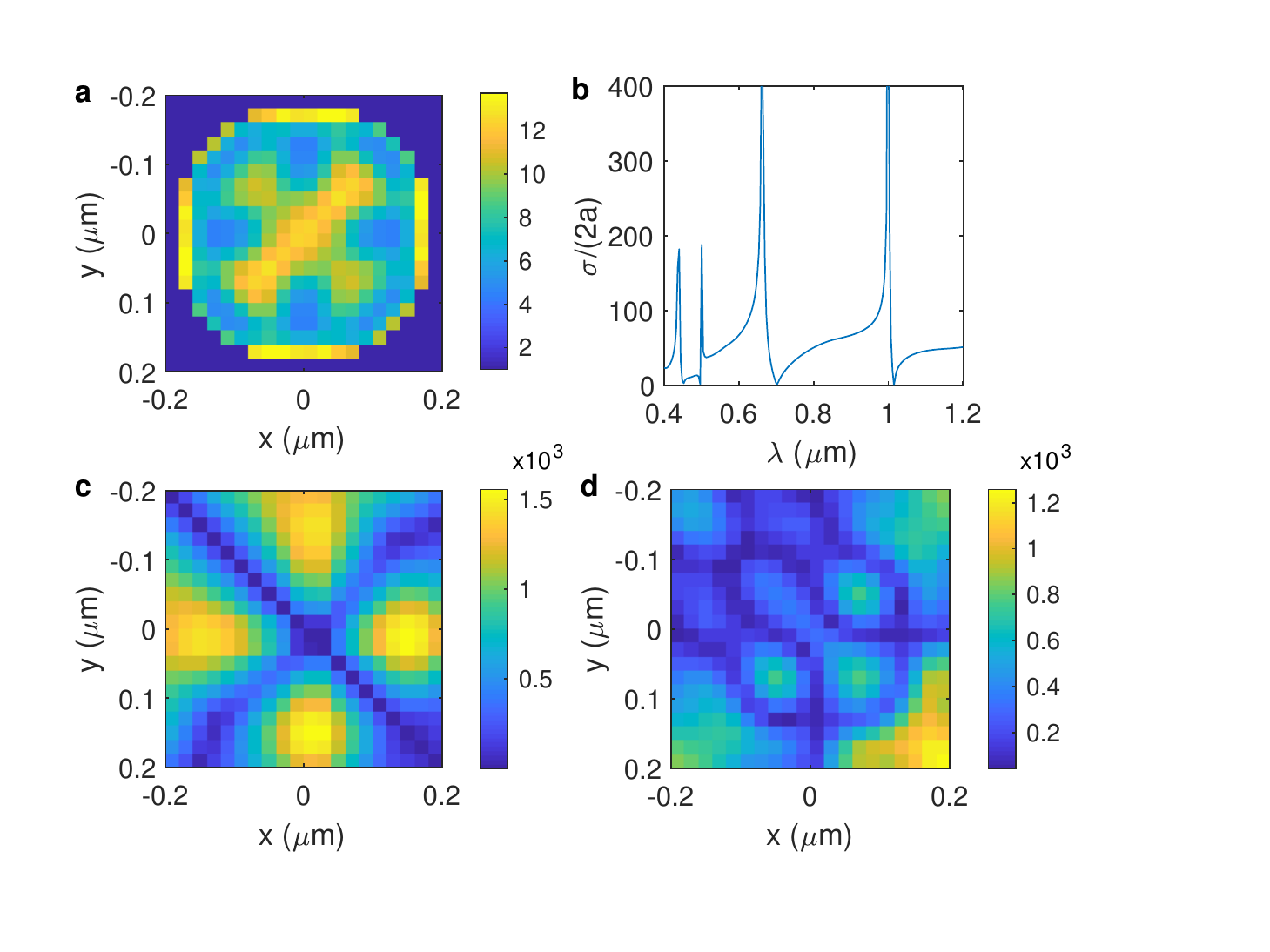}
\caption{\label{fig:epsart} Inverse design of a 0.4 $\mu$m structure with resonances at 1 $\mu$m and 0.5 $\mu$m. (a) Permittivity profile, (b) scattering cross section, and electric field at (c) 1 $\mu$m and (d) 0.5 $\mu$m.}
\end{figure} 

\section{Summary}

We developed a new inverse computational framework for designing photonic structures with multiple scattering resonances. In contrast with conventional inverse methods, our algorithm does not require solving Maxwell's equations for each material topology, and thus promises more efficient inverse computation. Our algorithm is based on optimizing a single constraint condition that defines the poles of the scattering matrix, which mathematically amounts to minimizing the determinant of the integral operator of the electric field with respect to the permittivity profile of the resonator. In this Letter, we developed the approach and implemented it to design several subwavelength resonant structures. First, we applied the method to show that a structure with a fixed form factor can be designed to resonate at different prescribed wavelengths, and the spectral variation of the cross sections are consistent with Fano lineshapes with $Q$s exceeding 500. Then we showed that the same structure can be designed to resonate at two prescribed wavelengths simultaneously. Such a structure may be used to enhanced second-harmonic generation. The compact form, computational efficiency, and generality suggest our method can be a useful inverse design tool for engineering resonators for enhanced light-matter interactions.

\section{acknowledgment}
The authors acknowledge support from DARPA under the NLM Program through Contract \# HR001118C0015.

\bibliography{bib}

\end{document}